\documentstyle[aps,epsf,twocolumn]{revtex}

\def\8{\infty}

\def\undertext#1{\vtop{\hbox{#1}\kern 1pt \hrule}}

\def\VEV#1{\left\langle\,#1\,\right\rangle}

\def\br{\\ \nonumber & &}

\def\be{\begin{equation}}
\def\ee{\end{equation}}
\def\bea{\begin{eqnarray} & &}
\def\eea{\end{eqnarray}}
\def\ct#1{\cite{#1}}
\def\rf#1{(\ref{#1})}

\begin{document}
\draft
\flushbottom
\twocolumn[
\hsize\textwidth\columnwidth\hsize\csname @twocolumnfalse\endcsname

\title{Conformal Algebras of 2D Disordered Systems}

\author {Victor Gurarie$^a$ and Andreas W.W. Ludwig$^b$}

\address{$^a$Institute for Theoretical Physics, University of California,
Santa Barbara CA 93106-4030}
\address{$^b$Department of Physics, University of California, Santa Barbara
CA 93106}

\date{\today}

\maketitle
\tightenlines
\widetext
\advance\leftskip by 57pt
\advance\rightskip by 57pt

\begin{abstract}
We discuss  the structure  of 2D conformal field theories (CFT)
at central charge $c=0$ describing critical disordered systems, 
polymers and percolation.
We construct a novel extension of the $c=0$ Virasoro algebra,
characterized by a number $b$ measuring the 
effective number of massless degrees of freedom,
and by  a logarithmic partner of the stress tensor.
It is argued to be present at a generic random critical point,
lacking super Kac-Moody, or other higher symmetries,
and is a tool to describe and classify such theories.
 Interestingly, this algebra
is not only  consistent with, but indeed naturally accommodates in general
an underlying global supersymmetry.  
Polymers and percolation realize this algebra. 
Unexpectedly, we find that the
$c=0$ Kac table of the degenerate fields contains two distinct theories
with $b=5/6$ and $b=-5/8$ which we conjecture  to  correspond to percolation and
polymers respectively. A given Kac-table  field 
can be  degenerate only in one of them.
Remarkably, we also find this algebra, and 
thereby an ensuing hidden  supersymmetry,  realized at general
 replica-averaged critical points, for which we derive an explicit formula
for $b$.

\end{abstract}
\vspace{1mm}

]
\narrowtext
\tightenlines

Critical behavior in systems with quenched disorder has remained
a major challenge in  condensed matter physics.
This area includes localization transitions of non-interacting
electrons,  a prominent example being the Integer Quantum Hall Effect
 plateau (IQHE) transition\ct{IQHE},  
but also various random statistical mechanics systems.
In two dimensions, powerful
techniques of Bethe Ansatz or Conformal Field Theory (CFT)
should lead to non-perturbative solutions.
Unfortunately this has not been the case so far.

Random critical points in two dimensions, when studied by replica
or supersymmetry (SUSY) methods, 
are believed to be described by  CFT\ct{BPZ} with central charge $c=0$.
In this paper we put forward a general algebra
describing  {\it supersymmetric}
disordered systems.
It extends the Virasoro $c=0$ algebra, and is
characterized by a parameter
$b$ similar to and extending the one found previously by one of us 
\ct{bTheor}.
The crucial novel feature is the appearance of logarithms
in the conformal symmetry generators themselves.
These then proliferate in almost all correlation functions.
Simpler disordered systems, including super Kac-Moody
current algebras, are rather special cases of this
where additional symmetries prevent these logarithmic
features from appearing. 
It is interesting to note that as soon as the
logarithmic features that we describe (such as e.g. Eq.\rf{AA})
 appear in {\it any $c=0$ conformal theory},
 SUSY can be accommodated automatically.


We show that this algebra is realized in polymers and percolation.
Even though a   great deal is known about properties of such 
geometrical models\ct{CardyHouches}  from their respective  mapping onto
 $q\to1$ and  $n\to 0$ limits of $q$-state Potts 
and $O(n)$ spin models,
the nature of the CFT at $q=1$ and $n=0$ is not so  well understood.
Indeed we show that 
at $q=1$ or $n=0$, where  both
models become manifestly supersymmetric\ct{Parisi,SQHE},
 these theories develop
precisely the logarithmic features of the kind described by our
formalism. These, apparently,  have been unnoticed in the past.
Many critical properties of polymers and percolation
have been described by $c=0$ Kac-table degenerate operators.
We show that the degeneracy condition of any Kac-table operator
fixes the value of $b$ of the theory uniquely.
We find, quite remarkably, that this allows for only two possible
values,  $b=5/6$ and $b=-5/8$, corresponding to two mutually
excluding sets of Kac-table primary operators being
degenerate.
 As a consequence, a given
$c=0$ Kac table operator  is degenerate only  for one choice,
not the other.
Therefore,  the form of correlation functions
as deduced from the $c=0$ Kac table degeneracy conditions
of a particular operator,
cannot be all valid simultaneously  for polymers and percolation.
We suggest that  $b=5/6$ corresponds to percolation,
and $b=-5/8$ to polymers.
This should be checked numerically.
For percolation this can be done using the
super spin-chain formulation of Ref. \ct{SQHE}.
A recent mapping
of the so-called Spin-Quantum Hall Effect (SQHE)  transition\ct{SQHESenthil}
onto percolation \ct{SQHE} may
be indication that the structure we describe is realized in a variety of
delocalization transitions in 2D. 
We believe that the algebra developed in this paper
is a tool to describe and classify generic
random critical behavior in 2D.



\begin{figure}[tbp]
\centerline{\epsfxsize=1in \epsfbox{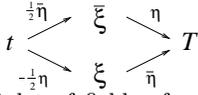}}
\caption {The multiplet of fields of conformal weight 2 for the group U$(1|1)$.
The arrows denote acting by the fermionic supersymmetry generators. The
stress tensor $T$ is invariant when acted upon by the supersymmetry
generators.}
\label{Pic1}
\end{figure}

{\it General Theory:} We start by considering
 a generic disordered system where the disorder average
can be performed using SUSY\ct{Efetov}.
It has been argued in \ct{bTheor} that
the stress tensor of such a  system is always  a member
of a  SUSY multiplet.  The number
of fields in this multiplet depends on the symmetry group of the system. 
Any such theory must at least be invariant
under a minimal $U(1|1)$ SUSY,   giving
rise to a 4-dimensional multiplet of stress tensors denoted  by $T$, $t$,
$\xi$ and $\bar \xi$ in \ct{bTheor}.
If the SUSY is larger, as it is for example the case for the SQHE\ct{SQHE} 
(where it is $ SU(2|1)$), this multiplet will contain more fields,
but the above four will always be contained therein.
It will suffice to consider those.
The action of the four
U$(1|1)$ generators,
two bosonic  ($J$ and $j$) and two fermionic  ($\eta$ and $\bar \eta$)
on the multiplet of stress tensors is
is schematically
depicted in Fig. \ref{Pic1}. 

In \ct{bTheor} a special role is played by the
`top' field $t$ of this multiplet, whose
correlation function with the stress
tensor $T$ was argued to be
\be
\label{Ttcorr}
\VEV{T(z) t(w)} = {b \over (z - w)^4}
\ee
where $b$ was a special parameter which counts the number of effective
degrees of freedom of the  disordered system. Different disordered
systems, all having central charge $c=0$,
 can be distinguished by different values of $b$. It was further
{ argued that the field $t(z)$ together with $T(z)$ should generate an
extension of the Virasoro algebra, via their operator product expansion (OPE).}
However, the most general form of this OPE was not established. Indeed, 
the OPE proposed in \ct{bTheor} was
\be
\label{Ttold}
T(z) t(0) = {b \over z^4} + {2 t(0) \over z^2} + {t'(0) \over z} + 
\dots.
\ee
Together with the OPE between the stress tensors
\be
\label{TT}
T(z) T(0) = {2 T(0) \over z^2} + {T'(0) \over z} + \dots.
\ee
(which simply reflects the fact the central charge of 
a disordered system is zero) the OPE \rf{Ttold} implies
\be
\label{ttold}
t(z) t(0) = {2 \kappa T(0) \over z^2} + {\kappa T'(0) \over z} + \dots,
\ee
where $\kappa$ is some parameter.

The OPEs \rf{Ttold}, \rf{TT} and \rf{ttold} are trivially realized in 
free theories, where disorder is `switched off', as is readily
verified by letting  $T=T_f+T_b$ and $t=T_f-T_b$. Here $T_{f/b}$ are
the stress tensors for the non-interacting  fermionic/bosonic 
parts.  Moreover, these OPEs are also realized in
certain super Kac-Moody algebras, including \ct{MCW}. 
However, these OPEs do not have enough structure to describe
generic random systems: by forming suitable
linear combination of $T$ and $t$, one simply obtains two
{\it commuting} Virasoro  algebras
with equal and opposite central charges. This shows that from the
point of view of conformal symmetry (i.e. Virasoro algebras),
a super Kac-Moody algebra is isomorphic to a non-random   system, albeit
with potentially non-trivial factors, replacing the free
theories with stress tensors $T_{f/b}$ above.

To rid ourselves from these shortcomings, we now propose
{ the following}
 generalization of the OPE  \rf{Ttold}, which is still
consistent with SUSY, but can accommodate
theories with a richer structure,
\be
\label{Tt}
T(z) t(0) = {b \over z^4} + {2 t(0) + \lambda T(0) \over z^2} + {t'(0)
\over z } + \dots.
\ee
with  some {\it non-vanishing} $\lambda$.
Indeed, by acting on this OPE with the symmetry generators, following
Fig. \ref{Pic1}, we can easily show that \rf{TT} is consistent with
\rf{Tt}. It is probably not possible to modify the OPE \rf{Tt} any further
while keeping it consistent with \rf{TT}. 
Since the OPE \rf{Ttold} describes only the  simple `factorized'
  theories discussed above,
 while the OPE \rf{Tt} appears to be  the most
general OPE we can write for a disordered system invariant under SUSY, 
we conjecture that the OPE \rf{Tt} is realized in a majority
of SUSY disordered systems, as opposed to special cases such as the ones
described by super Kac-Moody algebras or free field theories.  
Upon rescaling $t \rightarrow t/ \lambda $ ( thus: $b \to b/\lambda$)
we may set $\lambda=1$ in \rf{Tt}, a choice adopted from now on.
Observe also  that  a shift by $T$,  $t \to t + \gamma T$ preserves \rf{Tt}.


The OPE \rf{Tt} harbors many surprises. Its consequences
can be investigated 
with the help of conformal invariance alone,  {\it without} regard to SUSY. 
First of all, the form of the OPE \rf{Tt}
 implies that $t$ is a {\sl logarithmic operator}. 
Logarithmic operators, first introduced in \ct{Logs}, are responsible for
the appearance of logarithms in correlation functions directly at 
criticality,
without  violating scale or conformal invariance\ct{SaleurLogs}. 
The reason why  the OPE \rf{Tt} is compatible with,
and indeed responsible for, the appearance of logarithms, is because under
the action of the dilatation operator $L_0$, the field $t$ changes as
$L_0 t = 2 t + T$, that is, T is added to it. This behavior is mimicking
the behavior of the logarithms, which, under the change of scale
$z \rightarrow \lambda z$, change as $\log(z) \rightarrow \log(\lambda) +
\log(z)$. After Ref. \ct{Logs}, many authors worked out a great number
of properties of logarithmic operators \ct{Kogan}, whose results we will be using. 
However the logarithmic operator $t$ that we discuss here is special
in that it  is
a {\it logarithmic partner of the stress tensor} $T$ itself.
It is for this reason that its presence affects the entire theory
in a profound way. This distinguishes it
in a {\it crucial way} 
from other logarithmic operators discussed
elsewhere.

Next, { one can} show that conformal
invariance together with the OPE \rf{Tt} require that
\bea
\label{tt}
t(z) t(0) =  -{2 b \log(z) \over z^4}\br+ {t(0) \left[ 1- 4 \log(z) \right]
- T(0) \left[\log(z)+2 \log^2(z) \right] \over z^2} + \dots.
\eea
{ up to a certain ambiguity,
which we fixed by removing all 
non-logarithmic terms from 
 the two-point function $\VEV{t(z) t(0)}$, 
 by shifing with $T$}. 
This OPE replaces the trivial OPE \rf{ttold}. The OPE \rf{tt} contains 
logarithms which is the direct consequence of $t$ being logarithmic.

 Since the OPE \rf{tt} contains logarithms,
any correlation function of the form $\VEV{t(z) t(w) 
\dots }$ will not be single-valued. If we analytically continue 
 $z$ around $w$, a piece will be added to it. It 
possible to show that this piece will contain a term precisely  of the form 
$\VEV{\xi(z) \bar \xi(w) \dots}$ where $\xi$ and $\bar \xi$ are
(potentially identical)  primary 
operators with conformal weight 2 and a non-vanishing
two-point function. 
In cases of  $c=0$ theories with SUSY we clearly
want to match them with the fermionic
fields $\xi$ and $\bar \xi$ of the multiplet of Fig. \ref{Pic1}.
But  since our present discussion is more general, not assuming
any SUSY, these may be bosonic  {\it or} fermionic.
Let us normalize these operators  according to
\be
\label{xicorr}
\VEV{\xi(z) \bar \xi(w)} = {b \over 2 (z-w)^4},
\ee
The only OPE between $\xi$ and $\bar \xi$ compatible with the correlation function
\rf{xicorr}, with the OPE's \rf{TT},\rf{Tt},\rf{tt}, and  with
conformal invariance is
\be
\label{xx}
\xi(z) \bar \xi(0) = \alpha T(z) T(0) +
{b \over 2 z^4} + { t(0) +  T(0) \log(z) \over
z^2} + \dots.
\ee
where a term whose coefficient $\alpha$ is not fixed by conformal invariance
may be added. 
The remaining OPEs between the fields $\xi$, $\bar \xi$, $T$ and $t$ can also
be constructed by conformal invariance. For example,
\be
\label{tx}
t(z) \xi(0) = \beta T(z) \xi(0) 
- T(z) \xi(0) \log z + { \xi'(0) \over 2 z } + 
\dots.
\ee
Again the coefficient $\beta$ is not fixed by conformal invariance.
However, the associativity of the OPE in the
three-point function
$\VEV{t(z_1) \xi(z_2) {\bar \xi}(z_3)}$ fixes
$2 \alpha = \beta$. 

Now in a most remarkable way it is possible to show that the OPEs
\rf{TT}, \rf{Tt}, \rf{tt}, \rf{xx}, and \rf{tx}, worked out
from conformal invariance alone,  using \rf{TT}  and \rf{Tt} as a starting
point, are automatically covariant under the action of the supergroup, as depicted on
Fig. \ref{Pic1}. We do need to choose $\xi$ and $\bar \xi$ to be fermionic
fields and the coefficient $\beta=1/4$ in \rf{tx}, in order  to
make all these OPEs
covariant. But these were the free parameters left for us by conformal invariance.
Most of the terms fit into the multiplets by themselves, without any 
adjustments. 

The OPEs \rf{TT}, \rf{Tt}, \rf{tt}, \rf{xx}, and \rf{tx}
constitute  the first and central result of our paper.


One important comment is in order. The OPEs, such as \rf{tt} or \rf{xx},
contain logarithms and are  therefore not single-valued. One might think this
cannot be true, at least in a SUSY theory, since
 $t$ and $\xi$ are physical fields, obtained as the
``$zz$'' components of the
tensors $t_{\mu \nu}$ and $\xi_{\mu \nu}$ introduced in \ct{bTheor}. As
such, they have to be single-valued.  { As mentioned,  the
field $t$  (as obtained by supersymmetry), is defined up to
a shift $t \rightarrow t + \gamma T$.
 $\gamma$ can be regularization dependent},
and indeed, scale dependent at the critical point, because of $t$ being
logarithmic field. This is how the renormalized $t$ can become non 
single-valued without any contradictions. 

The most interesting consequence of our main result is as follows. Suppose
we have  a primary field $A$ in our system with conformal weight $h$. 
The OPE of this field with itself in,
say, the holomorphic sector, will contain
the contribution of the identity conformal block in the following way,
\be
\label{AA}
A(z) A(0) =  {1 \over z^{2 h}} \left[ 1 + {h \over b} \left\{ 
t(0) + \log(z)
T(0) \right\} z^2 + \dots \right], 
\ee
that is, it will contain logarithms. As a consequence, the
{ conformal block involving two 
 such  fields} will contain logarithms as well,
 as long as the field's conformal weight $h \not = 0$. This can be used as a test
if a given disordered system realizes the OPEs with logarithmic $t$
(see the examples below). 

Moreover, the above logic can be reversed. 
Once a given system exhibits the OPE \rf{AA}, we can argue that
it is supersymmetric. Indeed, once a primary operator obeys \rf{AA},
the field $t$ appears which has to obey \rf{Tt}. Once $t$ obeys
\rf{Tt}, the rest of the OPEs such as \rf{tt}, \rf{xx}, \rf{tx} follow
 from conformal invariance alone.
 Once these OPEs appear, they  are,
as mentioned above, 
automatically covariant under the SUSY action defined above!

{\it Percolation/Polymers:}
This ends our general discussion, and we will now  consider 
polymers and percolation (SQHE\ct{SQHE}) as particular examples.
Quite remarkably, we will see that they realize the OPE derived above
with a logarithmic $t$. 

\begin{figure}[tbp]
\centerline{\epsfxsize=1.2in \epsfbox{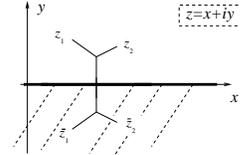}}
\caption {Points $z_1$ and $z_2$ in the bulk of the system and
their mirror image $\bar z_1$ and $\bar z_2$ across the boundary.}
\label{Pic}
\end{figure}

Consider for example the two-point function
of the (bulk)  energy operator 
$\epsilon$ 
in percolation of conformal weight $5/8$,
 in the presence of a conformally invariant boundary.
This is believed to be one of the degenerate
 Kac-table operators at $c=0$\ct{ChimZamo}.
This function  corresponds\ct{BoundaryCardy} to
a holomorphic four-point function without boundary,
with operators located at $z_1, z_2$ and their mirror images
at ${\bar z}_1, {\bar z}_2$, as in Fig. \ref{Pic}.
The latter can be found by solving the appropriate differential
equation, with solution 
\bea
\label{example}
\VEV{ \epsilon (z_1) \epsilon (z_2)}_{\rm bdr}=
{1 \over |z_1-z_2|^{5 \over 2} (1-x)^{5 \over 4}} \times \br 
\left[ (1-x)^2 F\left(
-{1 \over 2},
{3 \over 2};3;1 - x\right) + 
C \ x^2 F\left( -{1 \over 2},
{3 \over 2};3;x \right) \right]
\eea
where $x={(z_1-z_2)(\bar z_1- \bar z_2) \over (z_1-\bar z_2)(z_2-\bar z_1)}$,
$F$ is the hypergeometric function and $C$ 
depends on the boundary condition. 
As the points move away from the boundary ($x \rightarrow 0$),
the correlation function must  turn into the {\it bulk}  two-point function,
implying 
\bea
\label{expa}
{
(1-x)^2 F\left(
-{1 \over 2},
{3 \over 2};3;1-x\right)  + C \
x^2 F\left(
-{1 \over 2},
{3 \over 2};3;x \right) \over (1-x)^{5 \over 4}}
\propto \br \propto 1 + {15 \over 32}
x^2 \log(x) + \dots, \ \ \qquad  x \rightarrow 0
\eea
independent of $C$. 
This demonstrates that this boundary two-point function in percolation,
which is given by the above
 (chiral) four-point function, indeed contains the logarithms
which are the hallmark of our formalism.
The logarithm appearing here
is completely consistent with the OPE \rf{AA} if $b=5/6$. (The function
\rf{example} remains single-valued because $x$ is always
a negative real number). 

We learn two things about percolation from Eq.\rf{example}:
(i) Since the logarithmic $t$ operator is known to appear,
the theory possesses global SUSY.
 This follows from the above correlation function,
computed by exploiting only conformal symmetry.
( A manifestly  SUSY  invariant formulation of percolation
 was only very recently found \ct{SQHE}). 
(ii) the number of physical degrees of freedom $b=5/6$!

The (bulk)   energy operator $\epsilon$ (Eq. \rf{example})
is itself logarithmic. Indeed,
the SUSY formulation of Ref. \ct{SQHE}  showed that 
it lies in a multiplet isomorphic to that of the stress tensors.
Therefore, it also has a logarithmic partner, just like stress energy tensor,
whose {\sl two-point} function contains logarithms, in 
agreement with results that can be obtained from the
 `replica' method of Ref. \ct{RecentCardy}. 
However, in spite of the
presence of the logarithms in the
holomorphic and antiholomorphic sectors 
one can obtain a single-valued 4-point function in the bulk \ct{GurLud}.



For polymers we may consider  similarly  
the two-point function of the  bulk  energy operator,
of conformal weight $1/3$,  with a boundary
as in Eq.\rf{example}.
This operator is   believed to be Kac-degenerate \ct{ZamoPolymer}.
Explicit calculation as above  shows that this function
contains again logarithms of the same form as in \rf{expa}, except that now the
coefficient of the log-term is different.
Matching this coefficient with \rf{AA} yields $b=-5/8$. Actually,
it appears that
all the differential equations following from the $c=0$ Kac table
degeneracies  are consistent
with  a logarithmic $t$ and its OPEs,  if $b$ is chosen to be either $5/6$ or
$-5/8$ depending on which operator we are taking.

This is the second important result of our paper. 
The $c=0$ Kac table contains operators
corresponding  to different values of $b$ if we believe
that they are degenerate in the standard way and satisfy the standard differential
equations. Therefore, such operators cannot be degenerate in the same theory,
if we believe in the correlation function of  Eq. \rf{example}.
Take for example percolation:
 we cannot trust the degeneracy  equation for the operator $1/3$,
because it leads to $b=-5/8$ different from the value  found from
Eq.\rf{example}  above, in the same theory.
  One could try to say that
perhaps the same theory contains different fields $t_b$ and $t_{b'}$ which
appear in the OPE \rf{AA} for different primary fields with different
values of $b$. However, that would violate conformal invariance  (to say
nothing about supersymmetry). Indeed, in that case the correlation function
$\VEV{t_b \ t_{b'}}$ would be inconsistent with global conformal invariance.
So, one should treat the differential equations for the
correlation functions of polymers and percolation with care. 

{\it Algebraic Formulation:} 
 We can set up a completely algebraic way
to compute the value of $b$ without solving the differential
equations.  This is the first step towards a completely
algebraic reformulation of the
 OPEs \rf{TT}, \rf{Tt}, \rf{tt}, \rf{xx}.
A study of this algebra and classification of its
representations is likely to lead to a description of a
 quite  general class of new disordered critical points.


 Concentrate for now on the
 part of the algebra generated by
$T(z)$ and $t(z)$ only.
The OPE between $t(z)$ and any primary field $A(w)$ reads (by analogy
with \rf{tx})
\be
t(z) A(0) = - T(z) A(0) \log(z) + {{1 \over 2} A'(0) \over z}+ \dots,
\ee
where $\dots$ represent the regular terms in the 
Laurent-expansion (no logarithms).
This allows us to set up a mode expansion of the field $t$ when acting
on any  primary field $A$ as follows:
\be
\label{defl}
l_n A(0) = \oint dz z^{n+1} \left[ t(z) + T(z) \log(z) \right] A(0).
\ee
The argument of the integral is arranged in such a way
 that it is single-valued as $z$
goes around $0$. This definition leads to the commutation relations
between $l_n$ and $L_n$ in a standard way 
\be
\label{Ll}
[L_n,  l_m ] = {b \over 6} (n^3-n) \delta_{n+m,0}+(n-m) l_{n+m} + n L_{n+m},
\ee
This, together with the Virasoro commutation relations 
 allows us to check that
the conformal weight $5/8$ operator's null vector is primary 
with respect to  {\it both},   Virasoro ($L_n$)  and $l_n$ generators, 
{\it only} if $b=5/6$.
This is in agreement with the value of $b$ extracted from
 the (chiral)  four-point function computed earlier in this paper. 
After checking the operators degenerate down to the 6th level, we discovered
that the degeneracy of all the operators of the form $(1,q)$ with $q>2$ which
we were able to check required
$b=-5/8$, while all others we checked required $b=5/6$. It is natural
to conjecture that this is true of all Kac-table operators, not only those
degenerate down to the 6th level, but the mathematical proof of that,
based on \rf{Ll} and the Virasoro algebra, has not yet been obtained. 


{\it Replica Theories:}
Remarkably, we can  explicitly
identify the logarithmic field $t$ in 'replica`
theories,  where no SUSY is manifest at the outset. For example in random
bond ferromagnets,   a multiplicatively renormalizable partner
 ${\tilde T}^a$ of the stress tensor $T$ appears, transforming in the
$n-1$ dimensional representation of
the replica permutation group \ct{AWWL,RecentCardy}.
 ${\tilde T}^a$ is primary of dimension $\Delta(n)\to 2$
as $n, c \to 0$. We have  verified that in this limit
the field $(\tilde T_a+{1 \over n} T)$ is finite
and satisfies the OPE's required of $t$, with 
$-1/2b ={ \partial \Delta(n) \over  \partial  c }_{|c=0}$.
Hence we arrive at the striking conclusion that
the replica treatment of such random systems (allowing
for interactions) harbors a hidden SUSY.
For polymers and percolation, $n$ is replaced by 
a corresponding  multiplicity, and 
$\Delta(n)$ naturally coincides with the
dimension of $(3,1)$ or $(1,5)$ operators.

The authors are grateful to I. Gruzberg, N. Read,  H. Saleur, and
especially to J. Cardy, 
for useful discussions. 
One of us (V.G.) is
 supported by  the NSF grant PHY-94-07194.

\vfill

\end{document}